\documentclass[structabstract]{aa}  
\usepackage{graphicx}
\usepackage{txfonts}
\usepackage{natbib}
\begin{document}
   \title{The role of OH in the chemical evolution of protoplanetary disks\\
          I. The comet-forming region}

   \author{G. Chaparro Molano
          \and
          I. Kamp
          }

   \institute{Kapteyn Astronomical Institute, Postbus 800, 9747 AV Groningen, The Netherlands
             }

   \date{2011}

 \abstract{
Time-dependent gas-grain chemistry can help us understand the layered structure of species deposited onto the surface of grains during the lifetime of a protoplanetary disk. The history of trapping large quantities of carbon- and oxygen-bearing molecules onto the grains is especially significant for the formation of more complex (organic) molecules on the surface of grains. 
}{Among other processes, cosmic ray-induced UV photoprocesses can lead to the efficient formation of OH. Using a more accurate treatment of cosmic ray-gas interactions for disks, we obtain an increased cosmic ray-induced UV photon flux of $3.8\times10^5$ photons cm$^{-2}$s$^{-1}$ for a cosmic-ray ionization rate of H$_2$ value of 5$\times10^{-17}$ s$^{-1}$ (compared to previous estimates of 10$^4$ photons cm$^{-2}$s$^{-1}$ based on ISM dust properties). We explore the role of the enhanced OH abundance on the gas-grain chemistry in the midplane of the disk at 10 AU, which is a plausible location of comet formation. We focus on studying the formation/destruction pathways and timescales of the dominant chemical species.}{We solved the chemical rate equations based on a gas-grain chemical network and correcting for the enhanced cosmic ray-induced UV field. This field was estimated from an appropriate treatment of dust properties in a protoplanetary disk, as opposed to previous estimates that assume an ISM-like grain size distribution. We also explored the chemical effects of photodesorption of water ice into OH+H.}{Near the end of the disk's lifetime our chemical model yields H$_2$O, CO, CO$_2$ and CH$_4$ ice abundances at 10 AU (consistent with a midplane density of 10$^{10}$ cm$^{-3}$ and a temperature of $20$ K) that are compatible with measurements of the chemical composition of cometary bodies for a [C/O] ratio of 0.16. This comparison puts constraints on the physical conditions in which comets were formed.}{} 


   \maketitle
%

\section{Introduction}

Understanding the evolution of gas-grain chemistry and the role of the dust size distribution in protoplanetary disks is paramount to understanding their history. This history is imprinted in the remnants of the formation of the disks, namely asteroids and cometary bodies in our solar system. However, in-depth theoretical and laboratory studies of gas-grain chemical networks have only been researched in recent times \citep{aikawa,willacy,roberts,oberg,visser,wkt}. Chemical networks in protoplanetary disk-like conditions have not been as thoroughly studied as in molecular cloud-like conditions \citep{herbst,bates,umist}. For these reasons, we are interested in exploring chemical pathways in different regions of the disk, focusing on the role of OH, a highly reactive radical that can change the molecular composition of the gas when efficiently formed. To accomplish that, we study the chemical evolution and the chemical network in two regions of the disk midplane, including enhanced OH formation from cosmic rays and photodesorption of water ice.\\

Our goal is to study the time-dependent chemistry in the cold comet-formation zone, which in a T Tauri-type protoplanetary disk can be located at 10 AU \citep{cometjup}. We include in our chemical network the most important adsorption-desorption processes for H$_2$O, CO, CO$_2$ and CH$_4$ \citep{herbst,lab,oberg}, which link the gas species to the icy surface of grains. We also use a refined approach to calculating the effect of cosmic rays on both gas and grains \citep{leger,cecchi,shenvd}. Cosmic rays are important because they drive the chemistry in some regions of the disk via the ultraviolet (UV) field produced by their ionization of H$_2$ \citep{prasad,sdl,gldh}. This secondary UV field can photodissociate many chemical species and change the chemical balance, especially in regions of the disk that are opaque to stellar and interstellar radiation: the highly energetic cosmic rays can reach deep into the midplane of the disk. For this reason, the midplane of the disk can be considered as a cosmic-ray dominated region.\\

To analyze the chemical evolution, the local environment of the chemical system should resemble the conditions in a protoplanetary disk as much as possible. One of the main tools for building chemical networks in astronomy, the \textsc{Umist} database for astrochemistry \citep{umist}, lists some important chemical processes that have been calculated for molecular cloud-like conditions. The rates for some of these processes need to be recalculated for use in protoplanetary disk chemical networks, using the appropriate values for density, temperature, dust parameters, and radiation field. Cosmic ray-induced UV photo-dissociation is one example of the type of processes for which rates are calculated in the \textsc{Umist} literature using molecular cloud-like grain parameters and H$_2$ abundances \citep{sdl,gldh}. In protoplanetary disks, dust grains are larger than in molecular clouds due to aggregation and coagulation processes \citep{dalessio,natta,dominik}. Therefore, the overall grain surface and the local UV dust extinction will be smaller under protoplanetary disk conditions than under molecular cloud conditions. Here, we calculate the cosmic ray-induced UV (CRUV) photoprocess rates using the appropriate dust grain parameters.\\

In the work by Dalgarno et al. \citep[see][]{sdl,gld,gldh} on the interstellar medium and molecular clouds (on which the \textsc{Umist}06 \citep{umist} CRUV rates are based), the local extinction of cosmic ray-induced UV photons is dominated by dust grains; the gas contribution to the extinction is not considered to be important. However, in protoplanetary disks the gas composition can be very different from that of the interstellar medium or molecular clouds (e.g. high abundances of CH$_4$) which can significantly increase the gas opacity. It is then possible that in some regions of the disk CRUV photons can be absorbed much more efficiently by the gas than by the dust.\\

The general understanding of photodesorption is getting more detailed nowadays \citep{andersson, arasa}. When a photon hits an adsorbed molecule it can photodissociate it, instead of directly desorbing the ice species into the gas phase. The products of this dissociation can either recombine on the surface or desorb individually. In the case of water, molecular dynamics studies of a pure water ice layer that is hit by a UV photon show that a water molecule is often desorbed as OH+H instead of desorbing intact \citep{andersson,arasa}. The effect of this constant OH formation channel through grains may drastically alter the gas-phase chemical evolution.\\

We can safely neglect the gas opacity at distances of approximately 10 AU in our model, because most of the material is frozen on the dust grain surface, thereby suppressing the gas phase abundances of many molecules. However, as we get closer to the star, the temperature is high enough to keep molecules in the gas phase. At 1 AU the gas opacity is therefore important for the chemical evolution. Another factor to consider is that gas-phase processes involving line absorption will be affected by the local temperature conditions. This implies that CRUV photodissociation cross sections will have to be recalculated and integrated properly into the rate equations. This will be the subject of a subsequent paper.\\

The structure of this paper is the following. In Section \ref{dm} we explain how we use the results of the disk modeling code \textsc{ProDiMo} \citep{wkt} as input for our model. Next we explain the role of cosmic rays (Section \ref{trocr}), including an appropriate calculation (i.e. following protoplanetary disk grain parameters) of cosmic ray-induced UV photoprocesses in a low-gas environment. Section \ref{cre} consists of a comprehensive account of the gas-grain chemical model including adsorption-desorption mechanisms. Section \ref{r} deals with the setup for the chemical evolution model, including an estimate of the initial set of chemical abundances that apply for a protoplanetary disk. In Section \ref{r1} we explore the full chemical networks at the comet formation zone, comparing the chemical abundances with those measured in cometary bodies. In Section \ref{ewine} we discuss the chemical effects of H$_2$O ice photodesorbing into OH+H instead of desorbing intact. A discussion of our results in Section \ref{disc} is followed by out conclusions (Section \ref{conc}), in which we summarize the main results of this study.


\section{Disk Model}\label{dm}

\begin{figure}
\begin{center}
\includegraphics[scale=0.52]{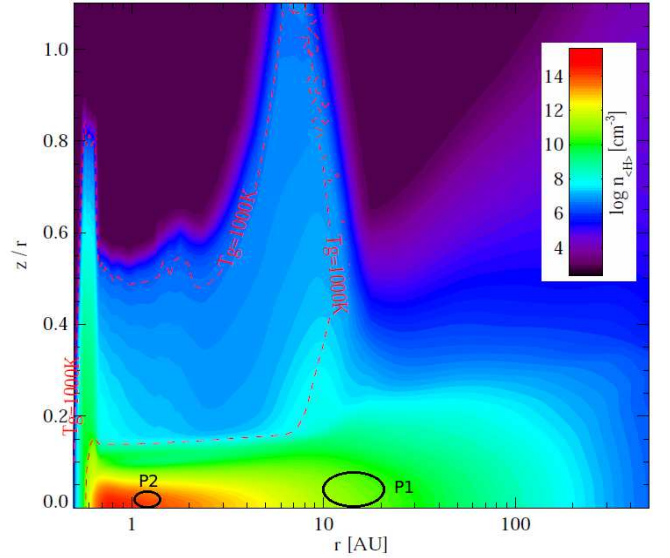}
\caption{Density structure (total hydrogen nuclei number density $n_\mathrm{\langle H\rangle}$) of a T Tauri disk model as function of radial distance from the star $r$ and the relative height $z/r$. Gas in thermal balance. The red dashed line shows the temperature boundary where $T_g=1000\ \mathrm{K}$. The regions under consideration for this paper are shown as black ovals (P1, and P2 for benchmarking). This figure is taken from the \textsc{ProDiMo} simulation \citep{wkt}.}\label{prod}
\end{center}
\end{figure}

For the sake of consistency, the disk structure used here is taken from the model of a passive disk around a T Tauri star obtained with the steady-state disk modeling code \textsc{ProDiMo} \citep{wkt}. This approach is valid because the accretion rate is negligible and transport processes occur on longer timescales than the ones modeled in this work. The position of the region under study within the disk structure (from the simulation) is illustrated in Fig. \ref{prod}. The location at 10 AU from the star corresponds to a likely location for comet formation, in which the density remains fairly high ($n_\mathrm{\langle H\rangle}$=10$^{10}$ cm$^{-3}$) if compared to molecular cloud conditions and the temperature is quite low ($T$=20 K). The CO and H$_2$O adsorption-desorption processes are most significant at such low temperatures\footnote{In those regions, the gas and grain temperatures are coupled, i.e. $T_\mathrm{gas}=T_\mathrm{dust}$ \citep{wkt}.}. Studying the change in chemical composition as a function of density while keeping a constant temperature is equivalent to moving in a direction perpendicular to the plane of the disk: disk models are vertically isothermal at high optical depth. Table \ref{tabpar} summarizes all relevant input parameters.\\

The grain size distribution in a quiescent protoplanetary disk is assumed to follow a power-law distribution $f(a)$$\sim$$ a^{-p}$ with $p$=3.5 and $a$ in the range 0.1-10 $\mu$m \citep{wkt}. This value of $p$ is also used for grains in molecular clouds \citep{draine}, but $a_\mathrm{min}$ and $a_\mathrm{max}$ are different, reflecting the overall smaller grain sizes in the interstellar medium (ISM): in the much denser protoplanetary disk environment aggregation processes and dust settling to the midplane produce larger sized grains on average. For example, \cite{dalessio} show the median Taurus SED and models it with an ISM grain size distribution and also with a distribution with larger grains on average. The latter is shown to fit much better. This grain growth causes a reduction of the average surface area of the grains, which limits their ability to absorb UV photons.\\

Dust settling in the midplane could lead to a lower gas-to-dust ratio than the typical value used here \citep{wkt}. This decreases dust extinction in the top layers in the disk (i.e. these layers become almost transparent), which allows the UV radiation to penetrate deeper towards the midplane. This means that the $A_V$=1 line is shifted to a lower height above the midplane. However, this effect can be compensated by the increased dust UV extinction in the midplane due to the enhanced dust-to-mass ratio. To isolate the effect of larger dust grains on the CR-induced UV field, here we stick to the grain parameters used in previous \textsc{ProDiMo} papers and keep the value 100 for the gas-to-dust mass ratio.\\

The local UV flux (in erg$\cdot$cm$^{-3}$) of the star and the ISM is frequently given in units of a Draine field \citep{draine,lee}. The strength of the UV field is then provided via the dimensionless quantity $\chi$ (which is 1 for the ISM),
\begin{equation}
 \chi=\left.\int_{91.2\,\mathrm{nm}}^{205\,\mathrm{nm}}\lambda u_\lambda\,d\lambda\ \middle/ \int_{91.2\,\mathrm{nm}}^{205\,\mathrm{nm}}\lambda u_\lambda^\mathrm{Draine}\,d\lambda\right.\ .
\end{equation}
The full 2D radiative transfer treatment in \textsc{ProDiMo} (which uses the same grain parameters, as stated in Table \ref{tabpar}) yields that for regions close to the midplane of the disk ($z/r$$<$$0.05$ and $0.7$$<$$r$$<$$10$ AU the disk is opaque to the stellar and interstellar UV photons and thus the local UV field strength is very low, $\chi\simeq0.001$ \citep{wkt}. Hence, UV photoreactions will only play a minor role here when compared to the effects of cosmic rays. \\

A cosmic ray-induced UV (CRUV) field is present in most regions of the disk \citep{prasad,cecchi}. This field can cause photodissociation reactions to take place at regions with high optical depth in the disk. However, CRUV photons are created locally and are assumed to be absorbed by the surrounding material (i.e. gas and dust grains) that is locally present. This means that the most appropriate treatment of CRUV photoreactions requires calculating the UV field created by cosmic-ray ionization of molecular hydrogen, accounting for dust grain surfaces as a possible sink for these UV photons by using the appropriate (local) grain properties, and also for the effects of gas opacity when the local conditions require. This will be discussed further in the following section.
\begin{table}
\begin{center}
\caption{Table of modeling parameters, following \cite{wkt}. The symbol \# identifies an ice-phase species.}
\label{tabpar}\renewcommand{\arraystretch}{1.3}
\begin{tabular}{lcc}\hline\hline
Parameter & Symbol & Value \\ \hline
Disk mass & $M_D$ & 0.01$M_\odot$\\ 
Dust-to-gas mass ratio & $\rho_d/\rho_g$ & 0.01\\ 
Minimum dust grain size & $a_\mathrm{min}$ & 0.1\,$\mu$m\\ 
Maximum dust grain size & $a_\mathrm{max}$ & 10\,$\mu$m\\ 
Grain size power-law index & $p$ & 3.5\\ 
Grain material mass density & $\rho_\mathrm{gmd}$ & 2.5\,g$\cdot$cm$^{-3}$\\ 
Grain mass density & $\rho_d$ & $2.16\times10^{-26}\cdot n_\mathrm{\langle H\rangle}\,\mathrm{g}$\\ 
Grain number density &$n_\mathrm{dust}$ & $4.6\times10^{-14}\cdot n_\mathrm{\langle H\rangle}$\\ \hline
Local UV field strength & $\chi$ & 10$^{-3}$\\ 
Dust opacity (UV, 100 nm) & $\kappa_\mathrm{UV}$ & 6.8$\times10^3$\,cm$^2\cdot$g$^{-1}$\\ 
Cosmic ray ionization rate (H$_2$) & $\zeta_\mathrm{H_2}$ & $5\times10^{-17}\,\mathrm{s^{-1}}$\\ \hline
Adsorption site area & $A_\mathrm{site}$ & $6.67\times10^{-16}$\,cm$^{2}$\\ 
Number of active layers & $N_\mathrm{Lay}$ & 2\\ 
Grain CR ``duty-cycle'' & $f(70\,\mathrm{K})$ & $3.16\times10^{-19}$\\ 
Grain albedo at 150 nm & $\omega$ & 0.57\\ \hline
CO Photodesorption yield & $Y_i$ (CO\#) & 2.7$\times10^{-3}$\\
H$_2$O Photodesorption yield & $Y_i$ (H$_2$O\#) & 1.3$\times10^{-3}$\\
CH$_4$ Photodesorption yield & $Y_i$ (CH$_4$\#) & $10^{-3}$\\
CO$_2$ Photodesorption yield & $Y_i$ (CO$_2$\#) & $10^{-3}$\\ \hline
CO Adsorption energy & $E_i^b$ (CO\#) & 960 K \\
H$_2$O Adsorption energy & $E_i^b$ (H$_2$O\#) & 4800 K \\ 
CH$_4$ Adsorption energy & $E_i^b$ (CH$_4$\#) & 1100 K \\
CO$_2$ Adsorption energy & $E_i^b$ (CO$_2$\#) & 2000 K \\\hline
\end{tabular}
\end{center}
\end{table}
\section{Cosmic rays}\label{trocr}

Cosmic rays play a very important role in the chemistry because they can pervade most regions of a protoplanetary disk. \cite{shenvd} have calculated that a cosmic-ray iron nucleus with an energy ranging from 20-1000 MeV will lose less than 1\% of its energy when passing through a dense molecular cloud. Their influence can be intense in highly obscured regions ($A_V\geq5$) where most stellar or interstellar UV radiation cannot penetrate \citep{roberts}. \cite{umebayashi} found that cosmic rays get effectively attenuated at a column density above $\Sigma$ $\sim$ 150 g cm$^{-2}$. In the particular disk model used here, the column density at 1 AU is not high enough to provide a significant extinction of cosmic rays. Choosing a steeper surface density distribution or a smaller outer radius could make the midplane opaque to cosmic rays at greater distances from the star. In \cite{leger} the total flux for cosmic-ray particles with energies higher than 0.02 GeV/nucleon in the ISM is calculated to be
\begin{equation}
 \Phi_{\mathrm{CR}}^{\mathrm{H}}=\int_{0.02}^\infty\phi_{\mathrm{H}}(\epsilon)\,d\epsilon=1.0\ \mathrm{CR\ cm}^{-2}\mathrm{ster}^{-1}\mathrm{s}^{-1}\ .
\end{equation}
The amount of cosmic raysthat are present can also be measured from the CR ionization rate of H$_2$, $\zeta_\mathrm{H_2}$. A recent work by \cite{indriolo} shows that column densities of H$_3^+$ (a tracer of H$_2$ ionization) vary in different molecular clouds. From their work, we pick a conservative value for $\zeta_\mathrm{H_2}$, which agrees with previous calculations obtained from measured cosmic ray spectra \citep{cecchi}
\begin{equation}\label{zeta}
  \zeta_\mathrm{H_2}=5\times10^{-17}\ \,\mathrm{s}^{-1}\ .
\end{equation} 

\subsection{Impact on chemistry}

Cosmic rays play an important role in molecular cloud chemistry \citep{klemperer,shenvd}: reactions with the CR-generated H$_3^+$ and He$^+$ are of special significance for CO/H$_2$O gas formation and destruction in the gas phase. Both H$_3^+$ and He$^+$ can dissociate CO and also SiO, which is important player in H$_2$O formation. He$^+$ is created from He+CR reactions, which happen at the same rate as H$_2$ ionization (see Eq. (\ref{zeta})) because they have approximately the same stopping power \citep{klemperer}. H$_3^+$ is created after H$_2$+CR collisions occur, which enable ionized molecular hydrogen to rapidly react with H$_2$
\begin{equation}
 \mathrm{H_2^++H_2\to H_3^++H}\ .
\end{equation}
As an important side effect, excited H$_2^+$ will also emit Werner and Lyman UV photons that can photodissociate otherwise stable species such as CH$_4$ \citep{prasad}. The presence of CR-induced UV photons can change the overall chemical balance through efficient ion-molecule chemical pathways.\\

For processes such as photodesorption, we need to calculate the local CRUV flux, $N_\mathrm{CU}$, while for CRUV photoprocesses we need to recalculate the efficiencies given the revised dust properties. The effects of gas opacity can be neglected at 10 AU, as most of the material is frozen on the grain surface, and gas abundances are low. The next two sections discuss this in more detail. 

\subsection{Cosmic ray-induced UV photon flux}\label{criuvp}

The effect of cosmic rays on molecular hydrogen is twofold: ionization and direct excitation \citep{cecchi}
\begin{equation}\label{cuch}
 \mathrm{CR}+\mathrm{H}_2\to e^-(30\mathrm{eV})+\mathrm{H}_2^+\ ,
\end{equation} 
\begin{equation}\label{cuch1}
 \mathrm{CR}+\mathrm{H}_2\to \mathrm{H}_2^*\ .
\end{equation} 
Fifty-five percent of all electrons produced in reaction (\ref{cuch}) will excite molecular hydrogen to excited electronic states \citep{sdl}:
\begin{equation}\label{cuch2}
 e^-+\mathrm{H}_2\to \mathrm{H}_2^*\ .
\end{equation} 
Excited electronic states of H$_2$  via (\ref{cuch1}) and (\ref{cuch2}) spontaneously decay to the $B$ $^1\Sigma_u^+$ and the $C$ $^1\Sigma_u$ states, which then decay to the ground electronic state $X\, ^1\Sigma_g^+$ and emit Lyman and Werner UV photons in the wavelength range $90-170$ nm:
\begin{equation}\label{cuch3}
 \mathrm{H}_2^*\to\mathrm{H}_2^{X\, ^1\Sigma_g^+}+\gamma_{\mathrm{UV}}\ .
\end{equation}
This is called the Prasad-Tarafdar mechanism \citep{prasad}.\\

We calculate the flux of locally generated cosmic ray-induced UV photons (in $\mathrm{cm^{-2}s^{-1}}$) based on the work of \cite{cecchi}:
\small
\begin{equation}
 N_\mathrm{CU}\simeq12\,500\ \frac{1}{1-\omega}\left(\frac{\zeta_\mathrm{H_2}}{5\times10^{-17}\,\mathrm{s}^{-1}}\right)\left(\frac{2\times10^{-21}\,\mathrm{cm^2}}{\sigma_\mathrm{\langle H\rangle}^\mathrm{UV}}\right)\left(\frac{n_\mathrm{H_2}/n_\mathrm{\langle H\rangle}}{0.5}\right)\ .
\end{equation}
\normalsize
Here $\omega=Q_\mathrm{sca}/(Q_\mathrm{sca}+Q_\mathrm{abs})$ is the grain albedo at $90-170$ nm, and $\sigma_\mathrm{\langle H\rangle}^\mathrm{UV}$ is the grain UV extinction cross section per hydrogen nucleus. The factor $n_\mathrm{H_2}/n_\mathrm{\langle H\rangle}$ is included because CRUV photons are generated by H$_2$ molecules, so their number should depend on the local H$_2$ abundance.\\

We calculate $\omega$ and $\sigma_\mathrm{\langle H\rangle}^\mathrm{UV}$ for the protoplanetary disk grain size distribution discussed in Section \ref{dm} using the appropriate dust opacity $\kappa_\mathrm{UV}$ (see Table \ref{tabpar}):
\begin{equation}\label{sigours}
\sigma_\mathrm{\langle H\rangle}^\mathrm{UV}=\kappa_\mathrm{UV}\frac{\rho_d}{n_\mathrm{\langle H\rangle}}=1.47\times10^{-22}\ \mathrm{cm^2}\ .
\end{equation} 
The \textsc{Umist} database uses a standard molecular cloud value for the UV extinction cross section \citep{sdl}
\begin{equation}\label{sigumist}
 \sigma_\mathrm{\langle H\rangle}^\mathrm{UV}=2\times10^{-21}\ \mathrm{cm^2}\ .
\end{equation} 
Such a divergence in $\sigma_\mathrm{\langle H\rangle}^\mathrm{UV}$ can be expected for a protoplanetary disk-like grain size distribution, since it favors large grains that do not provide significant extinction at short wavelengths when compared to a molecular cloud-like grain size distribution \citep{draine}. With this, it is possible to estimate a value for the CRUV photon flux under protoplanetary disk conditions, assuming that most of the H in the gas is stored in molecular hydrogen\footnote{This remains valid as long as molecular cloud-like abundances are used as initial conditions of the protoplanetary disk model.}:
\begin{equation}\label{cruvphot}
 N_\mathrm{CU}\simeq386\,000\mathrm{\ photons\ cm^{-2}s^{-1}}\ .
\end{equation} 
Here $N_\mathrm{CU}$ is 13.6 times more than the value implicitly used in \cite{sdl} (i.e. in the \textsc{Umist} database) and approximately 40 times higher than previous estimates for CRUV fields in molecular clouds \citep{shenvd}:
\begin{equation}
 N_\mathrm{CU}\simeq10\,000\mathrm{\ photons\ cm^{-2}s^{-1}}\ .
\end{equation} 
Comparing our value for $N_\mathrm{CU}$ with the stellar and interstellar UV field in the midplane (which is approximately 190$\,$000 photons cm$^{-2}$s$^{-1}$) and considering direct cosmic-ray ionization processes, it follows that cosmic-rays can be a major driver of the chemistry in the midplane of the disk. \cite{aikawacr} reached a similar conclusion based on the estimate of cosmic-ray extinction by \cite{umebayashi}.

\subsection{Cosmic ray-induced UV photodissociation}\label{crsigma}

The secondary field of cosmic ray-induced UV photons discussed in Section \ref{criuvp} has the same effect on the overall chemistry as stellar or interstellar UV radiation, via photodissociation/ionization of gas species and photodesorption of ice species. CR-induced photoprocesses are listed in the \textsc{Umist} database \citep{umist} using data from \cite{sdl} and \cite{gldh}. The cosmic-ray-induced photodissociation efficiency describes the competition for the locally generated UV photons between a species $i$ and the rest of the gas and dust grains:
\begin{equation}\label{fullint}
 \tilde{\gamma}_i=\int\frac{P(\nu)\sigma_i(\nu)}{\sigma_{\mathrm{tot}}(\nu)}\,d\nu\ .
\end{equation} 
Here $\sigma_i(\nu)$ is the photoprocess cross section \citep{lee}, $P(\nu)$ the H$_2$ line emission probability (Lyman and Werner lines in a de-excitation transition of H$_2$ from a CR-induced 30 eV electron excitation), and $\sigma_{\mathrm{tot}}(\nu)$ is the total extinction (of both gas and grain) cross section:
\begin{equation}
 \sigma_{\mathrm{tot}}(\nu)=\sigma_\mathrm{\langle H\rangle}^\mathrm{UV}(1-\omega)+\sum_{j}\frac{n_j}{n_\mathrm{\langle H\rangle}}\sigma_j(\nu)\ .
\end{equation}
Dust UV extinction is assumed to be larger than gas UV extinction in molecular cloud-like conditions \citep{sdl,cecchi}. However, for protoplanetary disk-type chemical abundances this only applies if the local gas-phase abundances are low, as is the case at 10 AU. Ignoring the contribution of the gas opacity, the efficiency is then written as
\begin{equation}
 \tilde{\gamma}_i=\int\frac{\sigma_i(\nu)P_i(\nu)}{\sigma_\mathrm{\langle H\rangle}^\mathrm{UV}(1-\omega)}\,d\nu=\frac{1}{1-\omega}\gamma_i\ .
\end{equation} 
The reaction rate coefficient is usually calculated by considering that CRUV photons are generated in a proportional amount to the CR ionization rate of H$_2$,
\begin{equation}
 k_{\mathrm{CU},i}=\zeta_\mathrm{H_2}\,\frac{1}{1-\omega}\,\gamma_i\ .
\end{equation} 
The previous expression ignores that CRUV photons are created from \emph{single} H$_2$ molecules, as opposed to hydrogen nuclei. For this reason the \textsc{Umist}06 database corrects the photodissociation efficiencies from \cite{gldh} by a factor 2, which is assumed to be the value for $n_\mathrm{\langle H\rangle}/n_\mathrm{H_2}$ in molecular clouds. For this reason we include the factor $n_\mathrm{H_2}/n_\mathrm{\langle H\rangle}$ explicitly in the CRUV photo-dissociation rate:
\begin{equation}\label{cruvrate}
 k_{\mathrm{CU},i}=2\,\frac{\zeta_\mathrm{H_2}}{1-\omega}\left(\frac{T_g}{300\ \mathrm{K}}\right)^{\beta}\frac{n_\mathrm{H_2}}{n_\mathrm{\langle H\rangle}}\,\gamma\ .
\end{equation} 
The exponent $\beta$=1.17 accounts for the fact that some species are photo-issociated by discrete line absorption \citep{umist,gld}. For all other species other than CO, this parameter is zero. Using our new value for $\sigma_\mathrm{\langle H\rangle}^\mathrm{UV}$ in Eq. (\ref{sigours}), CRUV photodissociation processes will be up to 13.5 times more efficient due to the reduced grain extinction\footnote{This factor is 13.5 when compared to the \textsc{Umist} rates, which are based on a standard molecular cloud-like dust UV extinction cross section. The factor 40 comes from the comparison to conservative estimates of the CRUV photon flux in molecular clouds, as in the previous section.}.

\section{Chemical model} \label{cre}

\begin{table}
\begin{center}
\caption{Table of chemical species in the chemical networks.}
\label{tabspe}\renewcommand{\arraystretch}{1.1}
\begin{tabular}{cc}\hline\hline
Type & Symbol  \\ \hline
Atoms & H, He, C, O, S, Si, Mg, Fe \\ \hline
Ions & He$^+$, Si$^+$, Fe$^+$, H$^-$, H$^+$, C$^+$, \\
 & O$^+$, S$^+$, Mg$^+$ \\ \hline
Molecules & H$_2$, H$_2$O, CH$_2$, HCO, SiO, CO$_2$, \\ 
 & SiH, CH$_3$, CH$_4$, OH, O$_2$, CO, CH  \\  \hline
Molecular & HCO$^+$, CH$_2^+$, H$_3^+$, SiH$^+$, SiO$^+$, \\
Ions &  CH$_4^+$, H$_3$O$^+$, H$_3$O$^+$, SiH$_2^+$, CH$_5^+$, \\ 
 &  CH$_3^+$, H$_2$O$^+$, SiOH$^+$, CH$^+$, H$_2^+$, \\ 
 & O$_2^+$, CO$^+$, OH$^+$, CO$_2^+$ \\ \hline
Ice species$^\mathit{a}$ &  H$_2$O$\#$, CO$\#$, CH$_4\#$, CO$_2\#$ \\ \hline
\end{tabular}\\
\textit{a}: \# stands for an ice-phase species.
\end{center}
\end{table}

The rate equations that describe the time-dependent chemistry (and chemical channels) of a gas-grain system comprise chemical reactions of the following types: 
\begin{itemize}
 \item gas-phase reactions: reactions between species that are present in the gas phase such as ion-molecule, recombination, charge transfer, neutral-neutral reactions, photo-dissociation and cosmic ray-induced processes.
 \item adsorption-desorption processes: reactions that take species from the gas phase into the solid phase (on the surface of the grain) or vice versa, usually driven by the local temperature, cosmic rays, and the local UV radiation field (stellar, interstellar and cosmic ray-induced).
 \item surface-surface reactions: reactions that occur on the surface of the grain, between species that are present in the solid phase.
\end{itemize}
Surface-surface reactions are beyond the current scope of this work, so only the first two are considered here. The 53 chemical species used here are listed in Table \ref{tabspe}.\\

The rate equation for the chemical network under study is a system of kinetically coupled differential equations. For a gas-phase species $i$, it is written as
\begin{equation}
 \frac{dn_i}{dt}=R_i^+-R_i^-\ .
\end{equation} 
Here $R_i^+$ is the rate that characterizes all processes that have the species $i$ as a product, and $R_i^-$ is the rate that describes all processes that decrease the abundance of the species $i$. For an ice species\footnote{The \# symbol identifies an ice-phase (frozen on a grain surface) species $i$.} $i\#$, the rate equation accounts for adsorption-desorption processes:
\begin{equation}
 \frac{dn_{i\#}}{dt}=k_i^\mathrm{a}n_i-k_i^\mathrm{d}n_{i\#}\ .
\end{equation} 
The four most important desorption processes considered here are thermal, photo-, cosmic ray-induced photo-, and direct cosmic-ray desorption:
\begin{equation}\label{despro}
 k_i^\mathrm{d}=k_i^\mathrm{d,th}+k_i^\mathrm{d,ph}+k_i^\mathrm{d,cr}+k_i^\mathrm{d,cu}\ .
\end{equation} 
It should be noted that we did not include X-ray desorption processes, which are very important near the inner rim, because the rates for these processes have not been determined beyond order-of-magnitude estimations \citep{walsh}.

\subsection{Adsorption}

Adsorption is the process that allows a species in the gas phase to adhere to the surface of a grain upon collision. Since the grain is assumed to be immersed in a well-stirred gas at a temperature $T_g$, a constant flux of gaseous molecules (each with mass $m_i$) is hitting the surface of the grain with an average thermal velocity of
\begin{equation}
 v_i^\mathrm{th}=\sqrt{\frac{kT_g}{2\pi m_i}}\ .
\end{equation}  
For a number density of dust grains $n_\mathrm{dust}$, a sticking probability $S=1$, and an average grain surface area of $4\pi\langle a^2\rangle$ the rate is
\begin{equation}
k_i^\mathrm{a}=4\pi \langle a^2\rangle\,S\,v_i^\mathrm{th}n_\mathrm{dust}\ .
\end{equation}
Adsorption becomes more efficient (i.e. acts on shorter timescales) as the gas density increases, given that more particles per volume are impinging on grains. The largest variation in the sticking parameter $S$ comes from the temperature-to-adsorption-energy ratio \citep[see][]{fegley}. In \cite{burke} a complete parameter-space exploration of the sticking probability yields that under protoplanetary disk-like conditions the chemical balance does not change significantly, which seem to be confirmed by experimental results in \cite{oberg}. Parameters for adsorption are given in Table \ref{tabpar}.

\subsection{Desorption}

Desorption processes counterbalance the complete depletion of species due to freeze-out. It should be noted that the details of the various desorption processes are not very well understood, are based on very few laboratory experiments, and use extreme simplifications \citep{roberts}. 

\subsubsection{Thermal desorption}

Thermal desorption is a process in which the temperature of the dust grain can cause some of the ice-phase species bound to its surface to evaporate. The rate at which this happens depends on a characteristic binding energy $E_i^b$:
\begin{equation}
k_i^\mathrm{d,th}=\nu_i\exp{\left(-\frac{E_i^b}{kT_d}\right)}\ .
\end{equation}
This Arrhenius-type expression for the thermal desorption rate is theoretical in nature, and the constant $\nu_i$ can be estimated from the vibrational frequency associated to the bond that holds the ice-phase species on the surface of the grain. This is based on the Polanyi-Wigner equation for a single desorption process, ignoring rotational degrees of freedom \citep{holloway,galwey}:
\begin{equation}
\nu_i=\sqrt{\frac{2n_{\mathrm{surf}}kE_i^b}{\pi^2m_i}}\ .
\end{equation}
Here $n_{\mathrm{surf}}=1/A_\mathrm{site}=1.5\times10^{15}$ cm$^{-2}$ is the number density of available adsorption sites per unit grain area \citep{surface}. Parameters for thermal desorption are given in Table \ref{tabpar}.

\subsubsection{Photodesorption}

Photodesorption occurs when a UV photon hits a specific adsorption site on the surface of the grain. The rate of this process can be written as
\begin{equation}\label{photor}
 k_i^\mathrm{d,ph}=\pi\langle a^2\rangle\frac{n_\mathrm{dust}}{n_\mathrm{act}}Y_i\,\chi F_{\mathrm{Draine}}\ .
\end{equation} 
This rate describes a Draine UV field\footnote{A value for the flux produced by this field  $F_{\mathrm{Draine}}\simeq2\times10^8$ cm$^{-2}$s$^{-1}$ is given in \cite{wkt}. The wavelength range for this field is given in Section \ref{dm} above.} impinging on the grain surface, and desorbing a species $i$. The photodesorption yield $Y_i$ is measured in the laboratory using a UV field comparable in strength to the Draine field \citep[see][]{oberg}. Given that the ice mantle on the grain surface is composed of many layers, the rate has to account for an incoming photon only being able to process a fraction of the icy molecules that are adsorbed onto the grain surface. This number of active sites on the grain surface is
\begin{equation}
 n_\mathrm{act}=4\pi\langle a^2\rangle n_\mathrm{dust}\,n_\mathrm{surf}N_\mathrm{Lay}\ .
\end{equation}  
Here $N_{\mathrm{Lay}}=2$ is the number of layers of ice that can be affected by an incoming UV photon. The expression for the photodesorption rate in Eqn. (\ref{photor}) can then be rewritten as
\begin{equation}\label{photo}
k_i^\mathrm{d,ph}=\frac{\chi F_{\mathrm{Draine}}}{4\,n_{\mathrm{surf}}N_{\mathrm{Lay}}}Y_i\ .
\end{equation}
Consequently, the rate for this process depends only on the amount of surface adsorption sites, the photodesorption yield, and the local flux of UV photons. This process can become important when the timescale for thermal desorption is large, such as for H$_2$O at temperatures below 120 K. Parameters for photodesorption are given in Table \ref{tabpar}.\\

Most species are assumed to photodesorb intact. However, this is not always the case for water, as reported in molecular dynamics studies by \cite{andersson} and \cite{arasa}. Most of the time (70\%) a water ice molecule that is hit by a UV photon will dissociate and subsequently desorb into an OH molecule and an H atom. This means that there is only a 30\% chance that the water molecule will desorb intact. It should be noted that this special case of water ice desorption does not change the photodesorption yield value. Here it is assumed that desorption occurs only for the first two layers of frozen material on the grain surface. When considering deeper layers, less than half of the water ice molecules desorb as water vapor. The effect on the overall chemical evolution of this special desorption case for water will be explored in Section \ref{ewine}. 
\begin{table}
\begin{center}
\caption{List of the most significant initial abundances of species used in the chemical networks.}
\label{tababu}\renewcommand{\arraystretch}{1.1}
\begin{tabular}{cc|cc}\hline\hline
\multicolumn{2}{c|}{Pre-molecular cloud} & \multicolumn{2}{c}{Molecular cloud} \\
\multicolumn{2}{c|}{(Atomic)} & \multicolumn{2}{c}{(Post-modeling)} \\ \hline
 Symbol & $\log(n_\mathrm{X}/n_\mathrm{\langle H\rangle})$ & Symbol & $\log(n_\mathrm{X}/n_\mathrm{\langle H\rangle})$ \\ \hline
 H & 0 & H$_2$ & -0.301 \\
 He & -1.125 & He & -1.125 \\
 O & -3.538  & CO & -3.939 \\
 C & -3.886  & H$_2$O$\#$ & -4.129 \\
 Si & -5.1 & O & -4.56 \\
 Fe & -5.367 & CO$_2\#$ & -4.575 \\ 
 Mg & -5.377 &  O$_2$ & -4.977 \\
 S & -5.721  &  SiO & -5.1 \\
 & & Fe & -5.367 \\
 & & Mg	& -5.377 \\
 & & H	& -5.57 \\
 & & H$_2$O & -5.798 $^\mathit{a}$ \\
 & & S & -5.721 \\ \hline 
\end{tabular}
\end{center}
\end{table}
\subsubsection{Cosmic ray-induced UV photodesorption}

This photodesorption process is identical to regular photodesorption, except for the source of UV photons. Therefore, by replacing the stellar and interstellar UV energy density $\chi F_\mathrm{Draine}$ in Eq. (\ref{photo}) by the appropriate CRUV photon flux for a protoplanetary disk $N_\mathrm{CU}$ in Eq. (\ref{cruvphot}), the cosmic ray-induced UV photodesorption rate becomes
\begin{equation}\label{cruvphd}
k_i^\mathrm{d,cu}=\frac{N_\mathrm{CU}}{4n_{\mathrm{surf}}N_{\mathrm{Lay}}}Y_i\ .
\end{equation}

\subsubsection{Direct cosmic ray desorption}

Grain heating by cosmic rays as described by \cite{herbst} happens when a cosmic ray passes through a disk and hits a dust grain, heating it up to an estimated temperature of 70 K. This causes species to thermally desorb from the grain as they would from thermal desoption at 70 K. Iron nuclei have been found to be the most important contributors to grain heating \citep{leger}. For an iron-to-hydrogen ratio of approximately [Fe/H]$\simeq$1.6$\times10^{-4}$, it is possible to estimate the rate at which cosmic rays hit an $a\simeq0.1$ $\mu$m grain:
\begin{equation}
 R_{\mathrm{CR}}=\Phi_{\mathrm{CR}}^{\mathrm{Fe}}\pi\langle a^2\rangle=3.16\times10^{-14}\,\mathrm{s}^{-1}\ .
\end{equation} 
The timescale for succesive CR-grain hits is then $\tau_\mathrm{CR}\simeq10^6$ yr. Given that the estimated cooling time for a silicate grain that goes from $T$=70 K to $T$=20 K is on the order of $10^{-5}$ s, we can find the ratio between the cooling and heating timescales\footnote{Calculation of this value requires knowledge of the specific heat obtained from indirect measurements, which means that it is highly dependent on $T$ and thus may not be valid for high ($>150$ K) dust temperatures.} with
\begin{equation}
 f(70\,\mathrm{K})=\frac{\tau_\mathrm{cool}}{\tau_\mathrm{CR}}=3.16\times10^{-19}\ .
\end{equation} 
This is called the ``duty-cycle'' of the grain heating by cosmic rays \citep{herbst}. The ``duty-cycle'' modulates the thermal desorption at $T$=70 K, scaled to the CR ionization rate of H$_2$
\begin{equation}\label{dcr}
k_i^\mathrm{d,cr}=f(70\,\mathrm{K})\,R_i^\mathrm{d,th}(70\,\mathrm{K})\,\zeta_\mathrm{H_2}\ .
\end{equation}

\section{Gas-grain chemistry model}\label{r}

The time-dependent chemical network used for this model is \verb"chem_compact", which is based on Milica Milosavlevic \& Inga Kamp's \verb"chemistry" code \citep{milica}. This code is a time-dependent solver of the gas-phase chemical rate equation and has been originally used to explore the role of shocks in protoplanetary disks. It uses the \textsc{Umist} database for astrochemistry \citep{umist}, which identifies all relevant chemical reactions for a set of chosen species. We use the ordinary differential equation (ODE) solver \verb"vode", which is specifically designed to solve stiff ODEs with strong and sudden time variations, such as the rate equation for a chemical network of any size \citep{vode}. This code has previously been benchmarked against steady state abundances, and in addition we make our own benchmark against \textsc{ProDiMo} \citep{wkt} steady state abundances in Section \ref{bench}.\\ 

Additionally, we include adsorption-desorption reactions for CO, CO$_2$, CH$_4$, and H$_2$O, and a new calculation of grain parameters for an appropriate treatment of CRUV photoprocesses for protoplanetary disk-like grain parameters (see Section \ref{trocr}).  We checked for the effects of incorporating other significant ice species such as O$_2$ and SiO ice to our network, and found no significant differences from our results. At 10 AU, O$_2$ and SiO ice are formed, but because adsorption is very efficient, it is necessary for them to be formed in the gas phase first. In future work, when we also consider surface reactions, we will incorporate adsorption for all gas-phase species.

\subsection{Initial conditions}

Assuming that the material has been processed before the formation of the disk, the initial abundances for the protoplanetary disk model were computed from molecular cloud-like conditions\footnote{This means that the molecular cloud values for the UV grain extinction cross section (Eq. \ref{sigumist}) were used in the CRUV photo-rates (Eqns. \ref{cruvrate} and \ref{cruvphd}).}. The chemical model was evolved from atomic conditions (see Table \ref{tababu}, left column) in a gas of density $n_\mathrm{\langle H\rangle}=10^6$ cm$^{-3}$, temperature $T=20$ K and $\chi=0.01$ for 10$^7$ years. The final chemical abundances for this run (see Table \ref{tababu}, right column) were used as input for the initial chemical abundances\footnote{The O$_2$ and H$_2$O abundances under molecular cloud-like conditions are inconsistent with current observations of the ISM \citep{o2inmc}. However, we found that formation of O$_2$ and H$_2$O in our protoplanetary disk model does not depend on their initial abundance.} in the different protoplanetary disk models studied here. A similar approach has been used by \cite{thi}.\\

We studied the chemical evolution under two different conditions for each point of interest: using a low (molecular cloud-like) and high (protoplanetary disk-like) value for the CRUV field. Thus we can see how the chemical abundances and pathways change as the protoplanetary disk-sized grains absorb less CRUV photons. 

\subsection{Chemistry benchmarking at 1 AU}\label{bench}

For the sake of consistency, we tested the solver against the steady-state chemistry in \textsc{ProDiMo} \citep{wkt}, by comparing the results at a distance of 1 AU from the star, near the midplane of the disk (P2 in Fig. \ref{prod}):  $T=80$ K, $n_\mathrm{\langle H\rangle}=10^{14}$ cm$^{-3}$ and $\chi=0.001$. For the purpose of matching the results of both models, we used a low (molecular cloud-like) value for the $\sigma_\mathrm{\langle H\rangle}^\mathrm{UV}$ and ignore the effects of CRUV desorption.\\

In the \textsc{ProDiMo}-simulated chemistry at 1 AU all the oxygen is trapped in H$_2$O ice ($n_\mathrm{H_2O\#}/n_\mathrm{\langle H\rangle}\simeq10^{-4}$), and all the carbon in methane gas ($n_\mathrm{CH_4}/n_\mathrm{\langle H\rangle}\simeq10^{-4}$), which implies a low abundance of gas-phase CO ($n_\mathrm{CO}/n_\mathrm{\langle H\rangle}\simeq10^{-6}$) and consequently a very low atomic oxygen abundance ($n_\mathrm{O}/n_\mathrm{\langle H\rangle}\simeq10^{-12}$). \verb"chem_compact" yields very similar results, although on a much larger timescale than the lifetime of the disk ($\tau\sim10^8$ yr). Therefore in this case, steady state abundances in the midplane have to be treated with caution. This has been pointed out by \cite{wkt}- see their Fig. 13.

\section{Chemistry in the comet-formation zone}\label{r1}

In the midplane of the disk at a distance of 10 AU from the central star, the density of our model is $n_\mathrm{\langle H\rangle}=10^{10}$ cm$^{-3}$ and the temperature is $T=20$ K, corresponding to the region P1 in Fig. \ref{prod}. Even though photo-desorption here is more efficient than thermal desorption, it acts at a timescale much longer than the disk's lifetime. These conditions are ideal for ice formation on grains, which means that carbon and oxygen will be trapped on the surface of grains and will not efficiently form gas-phase molecules. The chemical evolution in this region of the disk is plotted in Figs. \ref{excu10l} and \ref{excu10h}, corresponding to a low and a high CRUV flux, respectively.\\

From these figures it follows that while the local CRUV field does not radically change the chemical balance, it does have an effect. This will be discussed in Section \ref{creff}. It should also be noted that cosmic-ray ionization of H$_2$ and He has a very big impact on the chemistry, even though those rates are not affected by the dust grain parameters. The formation pathways that we describe in the following section apply to both low and high CRUV flux conditions.\\

\subsection{Chemical pathways}
\begin{figure}
\begin{center}
\includegraphics[angle=270, scale=0.7]{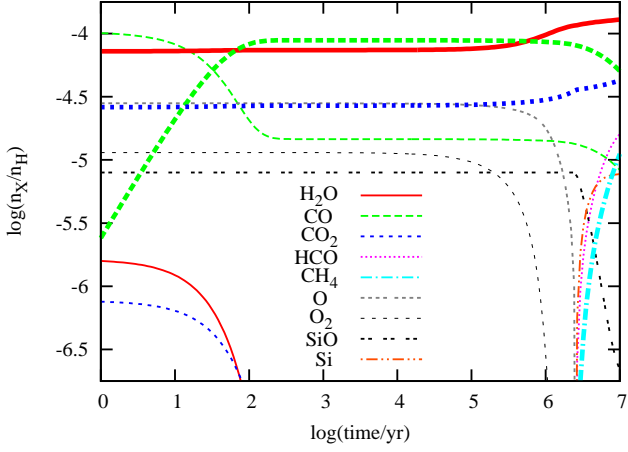}
\caption{Example of the time evolution of gas/ice abundances (thin/thick lines) using the standard ISM value for $\sigma_\mathrm{\langle H\rangle}^\mathrm{UV}=2\times10^{-21}\ \mathrm{cm^2}$. The disk parameters correspond to point P1 in Fig. \ref{prod} ($n_\mathrm{\langle H\rangle}=10^{10}\ \mathrm{cm}^{-3}$ and $T=20$ K).}\label{excu10l}
\end{center}
\end{figure}

The final abundances and time evolution of CO shows that it suffers depletion on a typical timescale of 10$^6$ yr. This suggests that H$_2$O is more efficiently formed than CO on long timescales, removing oxygen from the gas phase and trapping it into water ice. The main processes that create CO, H$_2$O and other related species are illustrated in Figs. \ref{chem1} and \ref{chem2}, where the gray pathways indicate CRUV photodissociation.\\

\begin{figure}
\begin{center}
\includegraphics[angle=270, scale=0.7]{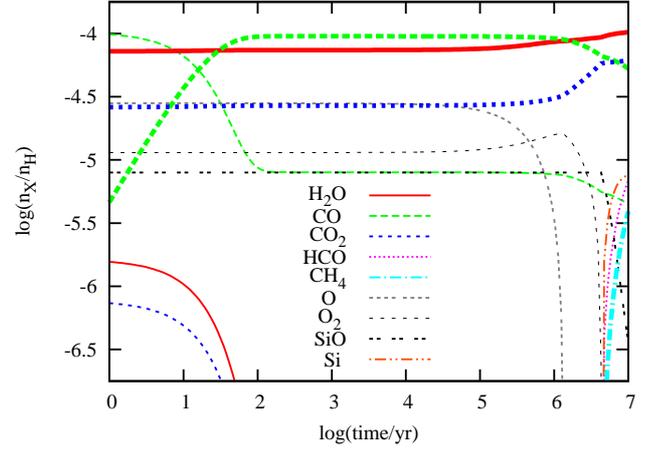}
\caption{Example of the time evolution of gas/ice abundances (thin/thick lines) using an appropriate $\sigma_\mathrm{\langle H\rangle}^\mathrm{UV}=1.5\times10^{-22}\ \mathrm{cm^2}$ for a protoplanetary disk. See text in Section \ref{creff} for a discussion of these results. The disk parameters correspond to point P1 in Fig. \ref{prod} ($n_\mathrm{\langle H\rangle}=10^{10}\ \mathrm{cm}^{-3}$ and $T=20$ K).}\label{excu10h}
\end{center}
\end{figure}

\begin{figure}
\begin{center}
\includegraphics[scale=0.55]{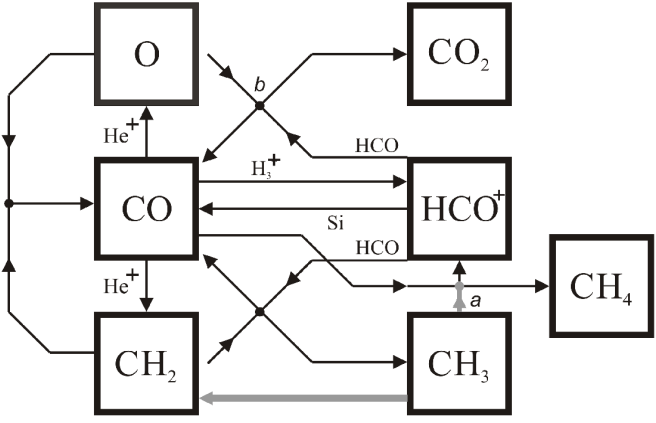}
\caption{Final ($10^{7}\mathrm{\ yr}$) chemical network for CO and CH$_4$ at P1 in Fig. \ref{prod} ($n_\mathrm{\langle H\rangle}=10^{10}\ \mathrm{cm^{-3}}$ and $T=20$ K), where the low temperature freezes CH$_4$ after its gas-phase formation at 10$^6$ yr, as seen in Fig. \ref{excu10h}. The lines represent reactions between species on the diagram, with arrows pointing at the product(s) of each reaction. Gray lines: Cosmic ray-induced photodissociation reactions. (Notes: \textit{a}. Via $\gamma_\mathrm{CRUV}$/H$_3^+\to$CH$_3^+$/CH$_4^+\to$CH$_5^+$. \textit{b}. HCO$^+$+O$\to$CO$_2$+H or  HCO$^+$+O$\to$CO+OH.)}\label{chem1}
\end{center}
\end{figure}

\begin{figure}
\begin{center}
\includegraphics[scale=0.55]{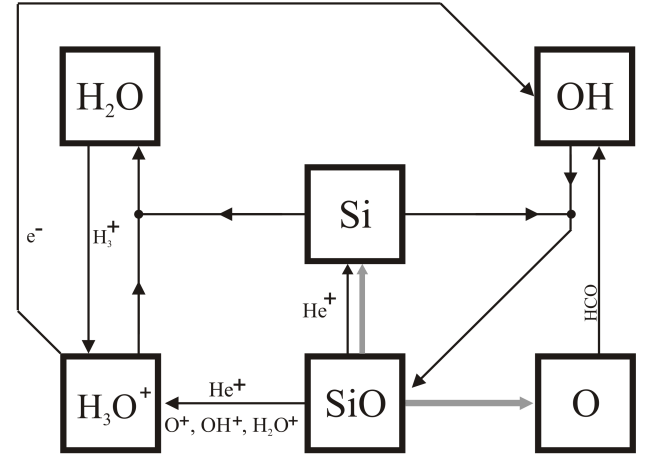}
\caption{Final ($10^{7}\mathrm{\ yr}$) chemical network for H$_2$O at P1 in Fig. \ref{prod} ($n_\mathrm{\langle H\rangle}=10^{10}\ \mathrm{cm^{-3}}$ and $T=20$ K). The lines represent reactions between species on the diagram, with arrows pointing at the product(s) of each reaction. Gray lines: Cosmic ray-induced photo-dissociation reactions.}\label{chem2}
\end{center}
\end{figure}

Figure \ref{excu10h} shows that after 100 yr, CO ice is the main carbon carrier. Excluding adsorption, the main reaction destroying gas-phase CO is
\begin{equation}\label{co-2}
 \mathrm{CO+He^+\to C^++O+He}\ .
\end{equation}
It should be remembered that He$^+$ is created by cosmic-ray ionization of He. C$^+$ created in this reaction will undergo charge exchange with other atoms (Si, Mg, Fe) to form atomic carbon, which then reacts very efficiently via radiative association with H$_2$ to form CH$_2$:
\begin{equation}
 \mathrm{C+H_2\to CH_2+\gamma}\ .
\end{equation} 
In turn, CH$_2$ reacts with atomic oxygen to again form CO:
\begin{equation}\label{co+1}
 \mathrm{CH_2+O\to CO+H_2}\ .
\end{equation} 
This reaction is very fast because atomic oxygen is steadily produced by He$^+$ dissociation of CO in reaction (\ref{co-2}). CO is also formed from HCO$^+$, either in the Si-reaction,
\begin{equation}
 \mathrm{HCO^++Si\to CO+SiH^+}\ ,
\end{equation}
or via HCO
\begin{equation}\label{hco+}
 \mathrm{HCO^++Fe,Mg\to HCO+Fe^+,Mg^+}\ ,
\end{equation}
\begin{equation}
 \mathrm{HCO+O\to CO+OH}\ ,
\end{equation}
an HCO$^+$ can itself be formed from CO via H$_3^+$
\begin{equation}\label{co-1}
 \mathrm{CO+H_3^+\to HCO^++H_2}\ .
\end{equation}
However, this HCO$^+$-CO feedback cycle is not closed because CO$_2$ can also be formed from HCO:
\begin{equation}\label{co2+1}
 \mathrm{HCO+O\to CO_2+H}\ .
\end{equation}
After formation, CO$_2$ is rapidly frozen, at the expense of a significant fraction of CO. This reduces the carbon available for CO ice formation.\\

CO$_2$ ice is more abundant in Fig. \ref{excu10h} than in Fig. \ref{excu10l} at the expense of water, HCO and CH$_4$. The long timescale for CH$_4$ ice formation is due to the chemical pathways that create CH$_4$ from CO. They are activated via CR-induced dissociation of CH$_3$, which acts on a timescale of approximately $8\times10^{6}$ yr. CH$_3$ is created from HCO and CH$_2$
\begin{equation}\label{co+3}
 \mathrm{CH_2+HCO\to CO+CH_3}\ .
\end{equation}  
The first and most important pathway starts from H$_3^+$-dissociation of CH$_3$. It should be noted that H$_3^+$ is created after cosmic-ray ionization of H$_2$:
\begin{equation}
 \mathrm{CH_3+H_3^+\to H_2+CH_4^+\left(+H_2\to CH_5^++H\right)}\ .
\end{equation}
The second pathway is started by CRUV photo-ionization of CH$_3$ (which is half as efficient as He$^+$ dissociation):
\begin{equation}
 \mathrm{CH_3+\gamma_{CU}\to e^-+CH_3^+\left(+H_2\to CH_5^+\right)}\ .
\end{equation} 
Thus, if CH$_3$ is efficiently dissociated, CH$_5^+$ will be efficiently formed. CH$_5^+$ is only one step away from forming CH$_4$:
\begin{equation}\label{co-4}
 \mathrm{CH_5^++CO\to HCO^++CH_4}\ .
\end{equation}
This reaction is more efficient at forming HCO$^+$ than reaction (\ref{co-1}). After CH$_4$ is formed, it is rapidly adsorbed onto the grain surface.\\

SiO forms H$_3$O$^+$ in reactions with He$^+$, which can form SiO again via OH. This feedback cycle is broken on the same timescale as the freeze-out of water, which causes the SiO abundance to decrease on long timescales (see Figs. \ref{excu10l} and \ref{excu10h}). The SiO/H$_2$O feedback cycle goes as follows:
\begin{equation}\label{si+1}
 \mathrm{SiO+He^+\to O^++Si+He}\ ,
\end{equation} 
\begin{equation}
 \mathrm{O^++H_2\to OH^++H}\ ,
\end{equation} 
\begin{equation}
 \mathrm{OH^++H_2\to H_2O^++H}\ ,
\end{equation} 
\begin{equation}
 \mathrm{H_2O^++H_2\to H_3O^++H}\ .
\end{equation}
Then, $\mathrm{H_3O^+}$ efficiently forms H$_2$O via dissociative recombinations \citep{bates,sdl}:
\begin{equation}\label{wat+si}
 \mathrm{H_3O^++e^-\to H+H_2O}\ .
\end{equation}
This reaction creates water vapor fairly efficiently on a timescale of 0.06 yr. However, there is another dissociative recombination reaction with Si, driven by the high abundance of atomic silicon formed in reaction (\ref{si+1}):
\begin{equation}\label{h2o+1}
 \mathrm{H_3O^++Si\to SiH^++H_2O}\ .
\end{equation}
The previous cycle continously generates water vapor that can rapidly be adsorbed onto the grain surface. This chain of reactions is very efficient because no other processes are creating O$^+$, OH$^+$ or H$_2$O$^+$. \\

On the other hand, water (in the gas phase) can go back to H$_3$O$^+$ via H$_3^+$
\begin{equation}\label{wat-1}
 \mathrm{H_2O+H_3^+\to H_3O^++H_2}\ .
\end{equation} 
The dissociative recombination of H$_3$O$^+$ and an electron can also create OH
 \begin{equation}
 \mathrm{H_3O^++e^-\to OH+H_2}\ .
\end{equation}
OH is a low-abundance but rapid catalyst for SiO formation
\begin{equation}
 \mathrm{OH+Si\to SiO+H}\ .
\end{equation} 
However, since OH forms SiO about ten times faster than SiO forms H$_3$O$^+$, SiO will be steadily formed within 10$^6$ yr, despite constant dissociation by CRUV photons. When we freeze all the water formed in reaction (\ref{h2o+1}), then the reaction (\ref{wat-1}) will be interrupted, and the OH-fueled SiO formation cycle is broken. Thus, even though CRUV photodissociation of water vapor into OH and H can favor SiO formation over water on long timescales, in neither case (low or high CRUV field) can we produce enough OH to keep high abundances of SiO in the gas phase after a few Myr, as seen in Figs. \ref{excu10l} and \ref{excu10h}. SiO is depleted after a few Myr because its formation pathway, reactions (\ref{si+1}) to (\ref{h2o+1}), is disrupted by freeze-out of water.

\subsection{The effect of CRUV enhancement}\label{creff}

Besides being important drivers for the later evolution of the chemistry, as seen in the previous section, CRUV photons affect different species in different ways. This is evident when comparing the chemistry that arises in the low and high CRUV flux environments, as seen in Figs. \ref{excu10l} and \ref{excu10h}. For instance, since CH$_3$ has an estimated higher cross section than CO and CH$_4$, it is more susceptible to CRUV photodissociation. Thus, when the CRUV flux is enhanced, the CH$_4$ formation pathways will be slowed down and its abundance will decrease. Also, even though there will more carbon available for CO formation, gas-phase CO will be more efficiently dissociated than CO$_2$. This shows in Fig. \ref{excu10h}, where we see the (final) CO$_2$ ice abundance increasing more (27\%) than the gas-phase CO (7\%) in the low CRUV flux case, in Fig. \ref{excu10l}.\\

This CO$_2$ ice enhancement comes at the expense of water, which decreases about 20\% in abundance because of CRUV photodissociation . This process enhances the OH abundance, and this shows how SiO can survive for a few more million years (as noted in the previous section). This shift in abundances is further proven by the long-term availability of atomic oxygen (formed in the photodissociation of CO and SiO), which in the high CRUV flux case is depleted a few hundred thousand years later than in the low CRUV flux case.

\begin{figure}
\begin{center}
\includegraphics[scale=0.55]{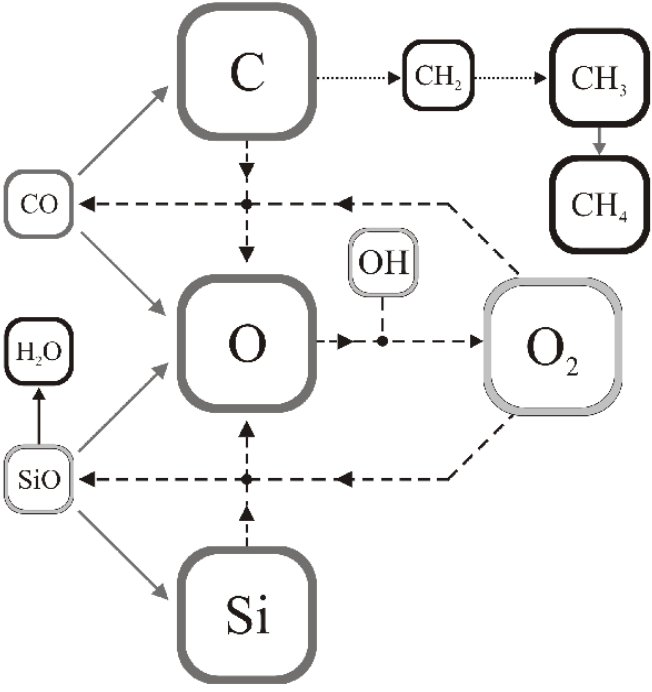}
\caption{Schematic chemical network for a protoplanetary disk, showing the competition of some chemical species (H$_2$O, CO, CH$_{3,4}$, SiO, O$_2$) for the carbon, silicon, and oxygen, depending on whether there is high OH formation or not. The lines represent reactions between species on the diagram, with arrows pointing at the product(s) of each reaction. The dotted branches correspond to a chemical network where OH is not being efficiently formed. The dashed branches replace the dotted pathways when OH is being efficiently created, thus taking away oxygen from H$_2$O to form CO, SiO, O$_2$, and atomic oxygen. The gray arrows represent CRUV photodissociation processes.}\label{chemf}
\end{center}
\end{figure}

\section{Alternative H$_2$O photodesorption mechanism}\label{ewine}

Molecular dynamics simulations by \cite{andersson} and \cite{arasa} show that the effect of UV photons impinging on water ice molecules is more destructive than previously thought. Before these studies, water molecules were believed to desorb intact into the gas phase after being expelled from the grain surface by a UV photon. However, water molecules are not desorbed instantaneously after the UV photon hits them: almost every single photon-water molecule interaction will result in photodissociation of the water molecule into OH and H. Most of the time (70\%), these byproducts will leave the surface in the gas phase. However, it is also possible (30\%) that the OH and H recombine on the surface, and the energy left over from this reaction will cause the newly formed water ice molecule to desorb into the gas phase.\\  

The main processes that create CO, H$_2$O, and other related species are summarized in Fig. \ref{chemf} schematically showing the chemical network depicted in Figs. \ref{chem1} (for CO, CO$_2$, and CH$_4$) and \ref{chem2} (for H$_2$O and SiO). They also show CRUV photodissociation processes and the new pathways arising from having an efficient OH formation mechanism.\\
\begin{figure}
\begin{center}
\includegraphics[scale=0.55]{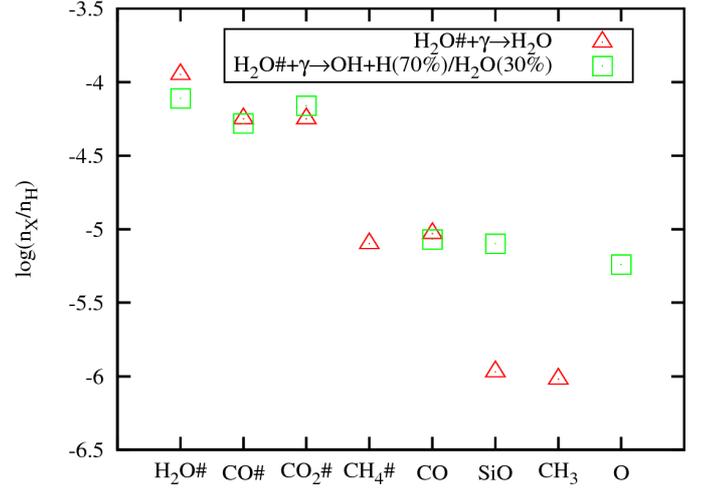}
\caption{Final ($10^7$ yr) abundances of significant species at the comet-forming region (P1 in Fig. \ref{prod}). Triangles: Water ice photodesorbs intact (Fig. \ref{excu10h}).  Squares: Water vapor is partially photodissociated into $\mathrm{OH+H}$. Plot obtained using an appropriate $\sigma_\mathrm{\langle H\rangle}^\mathrm{UV}$ for a protoplanetary disk and for a [C/O] ratio of 0.46.}\label{vdcold}
\end{center}
\end{figure}
When water ice photodesorbs intact at P1 (Fig. \ref{vdcold}, corresponding to the final abundances in Fig. \ref{excu10h}) most of the oxygen is in H$_2$O, CO, and CO$_2$ ice, and a fraction of carbon is in CH$_4$ ice (see \ref{chemf}). When water ice partially desorbs into OH and H (Fig. \ref{vdcold}) the efficient OH formation enables the oxygen in water to be stored elsewhere, such as in atomic oxygen and SiO. Now all the carbon is trapped in CO and CO$_2$ thanks to OH-driven CO formation (see Fig. \ref{chemf}).\\

The main chemical products of a high CRUV field present are atomic oxygen and, to a lesser degree, OH. Atomic oxygen, which is now very abundant will react very efficiently with desorbed OH to form O$_2$
\begin{equation}
 \mathrm{O+OH\to O_2+H}\ .
\end{equation}
This means that carbon and silicon photodissociated from CO and SiO will react with O$_2$ to form CO and SiO at approximately the same rate (while other carbon carriers like CH$_4$ are rapidly photo-dissociated):
\begin{equation}
 \mathrm{C+O_2\to CO+O}\ ,
\end{equation}
\begin{equation}\label{sif}
 \mathrm{Si+O_2\to SiO+O}\ .
\end{equation}
This feedback pathway arising from a highly efficient OH-forming mechanism enables the long-term gas-phase formation of O$_2$ and SiO that can later freeze on the dust grain surface. Now that carbon forms CO via O$_2$ and not via CH$_2$ as in reaction (\ref{co+1}), the byproducts CH$_3$ and CH$_4$ are not being efficiently formed, and CO becomes the main carbon carrier. The final chemical abundances resulting from having this special case of water desorption in P1 (see Fig. \ref{prod}) is depicted in Fig. \ref{vdcold}.

\section{Discussion}\label{disc}

In this work we focus on the timescales in which gas-grain chemical processes can take place in passive, nonaccreting disks. We thus assume that the density and temperature of the midplane will stay approximately constant during our simulation. We discuss the main results drawn from our study of the gas-grain chemical timescales in relation to observables such as the composition of cometary ices. 

\subsection{Implications for comet formation}\label{disc1}

Ice formation for different molecules takes place at different times because desorption processes act different timescales. This is evident in Fig. \ref{excu10h}, where CO ice is more abundant than water ice between 10$^2$ and 10$^6$ yrs. After this the ice content resembles what is observed in cometary ices\footnote{Even though ice ratios can vary among individual comets \citep[see][]{comet2}, the general composition $n_\mathrm{H_2O\#}>n_\mathrm{CO\#}>n_\mathrm{CO_2\#}>n_\mathrm{CH_4\#}$ is fairly consistent.} \citep{comet2,comet}, where the predominant form of ice is H$_2$O, followed by CO and CO$_2$ ice and, to a lesser degree, CH$_4$ ices. CH$_4$ ice is formed after a few million years, and only after water ice formation has become so efficient that it takes away most of the oxygen from CO and leaves the carbon free to form other molecules. This suggests that the CH$_4$ ice found in comets was formed very late in the disk evolution.\\ 

A comparison between measured cometary ice abundances and our resulting ice abundances is presented in Table \ref{tabcom}. Our resulting $n_\mathrm{CO\#}/n_\mathrm{CO_2\#}$ and $n_\mathrm{CH_4\#}/n_\mathrm{H_2O\#}$ ratios are very close to the cometary values, which means that the relative abundances of ice species are predicted by our model to some extent. Furthermore, when we attempted to increase $T$ (i.e. moving radially toward the star) or reduce $n_\mathrm{\langle H\rangle}$ (i.e. moving vertically away from the midplane) in order to decrease the adsorption rates, we ended up with a very different $n_\mathrm{CO\#}/n_\mathrm{CO_2\#}$ ratio, due to the high volatility of CO compared to CO$_2$. This means that the conditions at P1 are optimal for formation of comet-like ice abundances.\\

This conclusion is unique to the density-temperature combination rather than to the exact location in the midplane of the disk. If the temperature changes by 10 K, different ice abundances will vary in a nonuniform way due to the different binding energies of the chemical species; for example, at 10 K CH$_4$ ice is about 100 times less abundant because it cannot be formed efficiently. At 30 K thermal desorption of CO is so efficient that most of it stays in the gas phase. Also, if the density changes, the gas chemistry that drives the formation of ices via adsorption changes, so the particular ice structure will also change. At a density of 10$^{11}$ cm$^{-3}$, CO adsorption is so efficient that the gas-phase CH$_4$ formation processes are not efficient, and again lead to a CH$_4$ ice abundance that is 100 times less than in the 10 AU case. At a density of 10$^9$ cm$^{-3}$, CO$_2$ ice becomes twice as abundant as CO ice. Thus, a difference of an order of magnitude in density or 10 K in temperature will drastically change the ice composition.\\

The main reason for the discrepancy between the measurements compiled by \cite{comet} and our results stems from the fact that the carbon-to-oxygen ratio is much lower in observed cometary ices ($\sim$0.16) than in our model ($\sim$0.45). The reason behind this could be that some carbon is bound in dust that we consider here to be in the gas phase. In other words, as the sum of the carbon in the ices and in the gas phase is the total amount of carbon considered in our model, a fraction of it may be stored in refractory grain cores. Because of this we changed the [C/O] ratio in our model to the cometary value, which yielded the results in the right column of Table \ref{tabcom}. These results fit the cometary $n_\mathrm{CO\#}/n_\mathrm{H_2O\#}$ and $n_\mathrm{CO_2\#}/n_\mathrm{H_2O\#}$ ratios more closely. 

\section{Conclusions}\label{conc}
\begin{table}
\begin{center}
\caption{Ice ratios in comets compared to our results at two different epochs.}
\label{tabcom}\renewcommand{\arraystretch}{1.3}
\begin{tabular}{c|c|cc|c}\hline\hline
\multicolumn{1}{c|}{Ratio} & \multicolumn{1}{c|}{Cometary} & \multicolumn{2}{c|}{\texttt{c\_c}$^b$: [C/O]=0.45} & \multicolumn{1}{c}{\texttt{c\_c}$^c$: [C/O]=0.16} \\
 & measurements$^a$ & 10$^6$ yr & 10$^7$ yr & 10$^7$ yr\\ \hline
$n_\mathrm{CO_2\#}/n_\mathrm{CO\#}$ & $<$0.6 & 0.65 & 1 & 0.43\\ 
$n_\mathrm{CO\#}/n_\mathrm{H_2O\#}$ & $<$0.2 & 0.6 & 0.45 & 0.14 \\
$n_\mathrm{CO_2\#}/n_\mathrm{H_2O\#}$ & 0.02-0.12 & 0.39 & 0.6 & 0.06 \\
$n_\mathrm{CH_4\#}/n_\mathrm{H_2O\#}$ & 0.003-0.015 & 0.001 & 0.06 & $<$0.001 \\ \hline
\end{tabular}
\end{center}
\textit{a}: Abundances compiled from cometary measurements in \cite{comet2,comet}. \textit{b}: Ratios from our \texttt{chem\_compact} (\texttt{c\_c}) model using a [C/O] ratio of 0.45 (see Fig. \ref{excu10h}). \textit{a}: Ratios from our \texttt{chem\_compact} model using a (cometary) [C/O] ratio of 0.16.
\end{table}
Our time-dependent chemical network is compiled from the \textsc{Umist} dabase for astrochemistry \citep{umist}, adsorption-desorption processes for H$_2$O, CO, CO$_2$, and CH$_4$ \citep{wkt}, and an appropriate treatment of CRUV photoprocesses \citep{cecchi} for protoplanetary disks.\\

In the midplane of a protoplanetary disks, the steady formation of H$_2$O and CO and their relation to secondary oxygen and carbon carriers, such as CO$_2$, CH$_3$, CH$_4$ and SiO in the gas phase, are caused by recombination of ionized species. The presence of ionized material comes from an internal UV field, which in the dark, cold midplane reaches the floor level caused by cosmic ray interactions with the gas. Thus CRUV photons become the main driver of the chemistry. Locally generated UV photons can either photodissociate a molecule in the gas phase or impinge on the surface of a dust grain and photodesorb a frozen molecule. \\

By calculating grain parameters (such as the UV albedo and extinction cross section) for the grain size distribution appropriate for a protoplanetary disk, we find a CRUV photon flux of 380$\,000$ photon cm$^{-2}$s$^{-1}$, which is 40 times larger than conservative estimates for molecular clouds \citep{prasad,cecchi,shenvd,roberts}. Also, CRUV photodissociation processes can be up to 13.5 times more efficient in the midplane of a protoplanetary disk than in a similar molecular cloud-like environment. This can be curbed by the effects of gas opacity in warmer regions of the disk (closer to the star than 10 AU), which will be the subject of our next paper.\\

Cosmic ray-induced UV photons are responsible for the destruction of CH$_3$ (which favors CO formation) and formation of CH$_4$. Since CH$_4$ and CO$_2$ are more sensitive to CRUV photodissociation than CO, CO will mostly compete with H$_2$O as the most abundant oxygen-bearing species for the region of the disk that we probed here ($r$ $\sim$ 10 AU). Carbon and oxygen are efficiently trapped in ices in the midplane of the disk because the desorption timescales are longer than the lifetime of the disk.\\

The chemical abundances and their evolution change drastically whenever there is a mechanism that efficiently forms OH. A high formation rate of OH implies that O$_2$, SiO, and atomic oxygen will be efficiently formed, often at the expense of part of the oxygen in H$_2$O. Also, all the carbon will tend to be stored in CO instead of CH$_4$. \\

At 10 AU the only way to enhance OH formation is by photodesorbing water into OH+H. Even an enhanced CRUV field does not have any effect on OH formation because adsorption of gas molecules as soon as they are formed is very efficient at 20 K. When OH is not formed via photo-processing of water ice, we obtain ice-on-grain abundances and ratios that are comparable to those measured in comets \citep{comet2,comet}. One important instance is the formation of CH$_4$ ice, for which we obtained a timescale of a few Myr. This timescale is limited by (cosmic-ray generated) He$^+$ dissociation of CH$_3$.\\ 

The sensitivity of these ice ratios to temperature and density provide strong evidence that cometary ices must have formed under conditions similar to the ones used here, and most likely at late evolutionary stages (after a few Myr). This also means that the measured ice composition of comets precludes their formation in an OH-rich environment.

\begin{acknowledgements}
We would like to thank W.-F. Thi and P. Woitke for many helpful discussions on the effects of grain growth and on the physics behind CRUV processes, and M. Milosavlevic for providing us with the basis for the current \texttt{chem\_compact} code. Finally, we thank the anonymous referee and the A\&A Editor Malcolm Walmsley for helping us clarify important aspects of this work. 
\end{acknowledgements}

\bibliographystyle{aa}
\bibliography{bib}

\end{document}